\documentclass[oneside,12pt]{article}
\usepackage{amsmath}
\usepackage{amsfonts}
\usepackage{amssymb}
\usepackage{graphicx}

\begin{document}

\title{{\Large Cosmological parametrizations and their scalar field
descendents }}
\author{S. K. J. Pacif$^{1}$, K. Myrzakulov$^{2}$, R. Myrzakulov$^{3}$, \\
$^{1}$\textit{Centre for Theoretical Physics,}\\
\textit{\ \ Jamia Millia Islamia,}\\
\textit{\ \ New Delhi 110025, India}\\
\textit{\ }$^{2,\text{ }3}$\textit{Eurasian International Center for
Theoretical Physics,}\\
\ \ \ \textit{Department of General and Theoretical Physics,}\\
\textit{\ \ \ Eurasian National University, Astana 010008, Kazakhstan}\\
\textit{\ \ \ shibesh.math@gmail.com}$^{1}$\textit{, kairatmyrzakul@gmail.com%
}$^{2}$,\\
\ \ \ \textit{rmyrzakulov}@gmail.com$^{3}$}
\maketitle

\begin{abstract}
We reconstruct the field potentials in case of (non)phantom fields for
different models resulting from parametrization of $q(t)$, $a(t)$ or $H(t)$.
In addition we carry out similar procedure for tachyonic field. We also
discuss specific form of parametrization for reconstruction of scalar field
potential.
\end{abstract}


\section{Introduction}

\qquad The revolution in observational cosmology during the past two decades
has provided sufficient evidence for late time acceleration of the Universe 
{\small \cite{SCP, HZTEAM, HZ1, HZ2, HZ3, SCP1, SCP2, SCP3}}. This
phenomenon can be explained in several ways such as by incorporation of an
extra term in the right hand side of Einstein's field equations or by
modifying the left hand side of the field equations. In general relativity
the concept of \textit{dark energy} seems to be more relevant to the
observed accelerated expansion of the Universe. In this framework, dark
energy constitutes nearly 73\% of the total energy budget of the Universe
along with other components the - dark matter (23 percent) and the baryonic
matter (4 percent). However, important questions concerning the nature of
dark energy, its interaction with other material components in the Universe,
yet remain to be answered. A large number of candidates for dark energy
including cosmological constant have been proposed in the recent years (see
the review articles {\small \cite{SamiReview1, SamiReview2, SamiReview3,
SamiReview4, SamiReview5, SamiReview6, Yoo, ODINT-DE}}). Phenomenologically
quintessence field {\small \cite{quint1, quint2, quint3, quint4}} with
standard kinetic term and minimally coupled to gravity can be considered as
a very good candidate for dark energy. In slow roll approximation (potential
dominated scalar field i.e.$\frac{\dot{\phi}^{2}}{2}<<V(\phi )$), it can
also act as a cosmological constant. The scalar field with the wrong sign in
the kinetic term, dubbed phantom {\small \cite{phant1, phant2, phant3,
phantsami, phant4, phant5, Elizalde}} is also allowed observationally. There
are other scalar field models relevant to dark energy namely, quintom 
{\small \cite{quintom1, quintom2, quintom3, quintom4}}, k-essense {\small 
\cite{kessen1, kessen2}}, tachyon {\small \cite{tachy1, tachy2, tachy3,
tachy4, tachy5, tachy6}}, light mass Galileons {\small \cite{galileon1,
galileon2, galileon3, galileon4, galileon5, galileon6}}, chameleon {\small 
\cite{chamel1, chamel2, chamel3}} etc. There is plethora of field potentials
that can describe the smooth transition from deceleration to acceleration.
In this context various canonical as well as non canonical scalar field
potentials (e.g. exponential potential, flat potential, linear potential,
quadratic potential etc.) for different fields have been proposed that can
lead to different theoretical and observational consequences.

On the other hand the inclusion of one more component (dark energy) into the
evolution equations in the form of scalar field adds an extra degree of
freedom. And for a unique solution, one requires a constrain equation. This
can be achieved, in particular, by parameterizing the deceleration parameter 
$q(t)$, Hubble parameter $H(t)$, the equation of state parameter $w(t)$ or
the scale factor $a(t)$ ( for a recent review on various parametrization one
can see {\small \cite{SKJP}}). An interesting article {\small \cite%
{samireconstruct}} can be found in literature wherein tachyonic potential is
reconstructed on the FRW brane. There are other reconstructions of scalar
field potentials describing the late-time acceleration of the Universe e.g.
reconstructions of scalar field potential to unify early-time and late-time
Universe based on phantom cosmology {\small \cite{nojiri, capoz}},
reconstruction of scalar field potential in light of SN data {\small \cite%
{Li}}, reconstruction of phantom scalar potentials in two-field cosmological
models \cite{Andrianov}, holographic reconstruction of scalar field dark
energy models {\small \cite{surajit} }and many more. In this paper,
following {\small \cite{samireconstruct, nojiri}}, we reconstruct the scalar
field potentials for models obtained by various parametrization of $q(t)$, $%
a(t)$ or $H(t)$ in case of quintessence, phantom and tachyonic fields.

\section{Scalar field potentials for quintessence and phantom field}

We consider an action describing a general scalar field $\phi $ as

\begin{equation}
S=\int d^{4}x\sqrt{-g}\left\{ \frac{M_{p}^{2}}{2}R-\frac{1}{2}\omega
\partial _{\mu }\phi \partial ^{\mu }\phi -V(\phi )+L_{Matter}\right\} \text{%
,}  \label{A}
\end{equation}%
where $\omega =+1$ or $-1$ for quintessence and phantom field respectively
and $V(\phi )$ is the potential function for the scalar field. In the flat
FRW background the energy density $\rho _{\phi }$ and pressure $p_{\phi }$
of the scalar field can be written as

\begin{equation}
\rho _{\phi }=\frac{1}{2}\omega \dot{\phi}^{2}+V\left( \phi \right) \text{,}
\label{a}
\end{equation}

\begin{equation}
p_{\phi }=\frac{1}{2}\omega \dot{\phi}^{2}-V\left( \phi \right) \text{.}
\label{b}
\end{equation}

From equations (\ref{a}) and (\ref{b}) we may obtain

\begin{equation}
V(\phi )=\frac{1}{2}\left( \rho _{\phi }-p_{\phi }\right)  \label{d}
\end{equation}%
and

\begin{equation}
\omega \phi (t)=\int \left( \rho _{\phi }+p_{\phi }\right) ^{\frac{1}{2}%
}dt+\phi _{i}\text{ , where }\phi _{i}\text{ is a constant of integration.}
\label{e}
\end{equation}

The effective energy density and pressure can be written as

\begin{equation}
\rho _{eff}=\rho _{\phi }+\sum \rho _{i}\text{ and }p_{eff}=p_{\phi }+\sum
p_{i}\text{ ,}  \label{f}
\end{equation}%
where $\rho _{i}$ and $p_{i}$ are the energy densities and pressures of all
relativistic and non-relativistic components of the Universe. Using the
perfect fluid equation of state $p_{i}=w_{i}\rho _{i}$ ($0\leqslant
w_{i}\leqslant 1$) for the matter fields and substituting (\ref{f}) in (\ref%
{d}) and (\ref{e}), we may obtain the expressions

\begin{equation}
V(\phi )=\frac{1}{2}\left[ \left( 1-w_{eff}\right) \rho _{eff}-\left(
1-w_{i}\right) \sum \rho _{i}\right]  \label{g}
\end{equation}

and

\begin{equation}
\omega \phi (t)=\int \left[ \left( 1+w_{eff}\right) \rho _{eff}-\left(
1+w_{i}\right) \sum \rho _{i}\right] ^{\frac{1}{2}}dt+\phi _{i}\text{.}
\label{h}
\end{equation}%
where $w_{eff}=\frac{p_{eff}}{\rho _{eff}}$ is the effective equation of
state parameter. For flat ($k=0$) case, Friedmann equations reduce to

\begin{equation}
\rho _{eff}=3M_{p}^{2}H^{2}\text{,}  \label{i}
\end{equation}

\begin{equation}
p_{eff}=-M_{p}^{2}\left( 3H^{2}+2\dot{H}\right) \text{.}  \label{j}
\end{equation}

Observations suggest that the dominant constituents in the Universe are dark
energy and cold dark matter. So, considering a two fluid Universe (dark
energy and cold dark matter), equations (\ref{g}) and (\ref{h}) reduce to

\begin{equation}
V(\phi )=\frac{1}{2}\left[ \left( 1-w_{eff}\right) \rho _{eff}-\rho _{m}%
\right]  \label{k1}
\end{equation}

and

\begin{equation}
\omega \phi (t)=\int \left[ \left( 1+w_{eff}\right) \rho _{eff}-\rho _{m}%
\right] ^{\frac{1}{2}}dt+\phi _{i}\text{.}  \label{k2}
\end{equation}%
\qquad

Furthermore, if we assume the minimal interaction between matter and the
scalar field then from the conservation equation, we have $\ \dot{\rho}%
_{\phi }+3H\left( 1+w_{\phi }\right) \rho _{\phi }=0$ and 
\begin{equation}
\dot{\rho}_{m}+3H\left( 1+w_{m}\right) \rho _{m}=0\text{,}  \label{c}
\end{equation}%
which yields $\rho _{m}=\rho _{0}a^{-3}$, where $\rho _{0}$ is a constant of
integration and is generally attributed to present value of matter energy
density. Here and afterwards a suffix `$0$' for any variable refers to
present value of the concerned quantity. Hence, the potential for the scalar
field can be written as

\begin{equation}
V(\phi )=\frac{1}{2}\left[ \left( 1-w_{eff}\right) \rho _{eff}-\rho
_{0}a^{-3}\right] \text{,}  \label{G}
\end{equation}%
together with the expression of the scalar function%
\begin{equation}
\omega \phi (t)=\int \left[ \left( 1+w_{eff}\right) \rho _{eff}-\rho
_{0}a^{-3}\right] ^{\frac{1}{2}}dt+\phi _{i}\text{.}  \label{H}
\end{equation}%
From the two Friedmann equations (\ref{g}) and (\ref{h}), it is easy to
derive 
\begin{equation}
w_{eff}=-1-\frac{2}{3}\frac{\dot{H}}{H^{2}}\text{.}  \label{k}
\end{equation}%
which can also be represented as

\begin{equation}
w_{eff}=-\frac{1}{3}+\frac{2}{3}q=-\frac{1}{3}-\frac{2}{3}\frac{a\ddot{a}}{%
\dot{a}^{2}}\text{.}  \label{l}
\end{equation}%
\qquad \qquad

We can observe that, for any parametrization of the parameters $q(t)$, $H(t)$
or $a(t)$, all the quantities $\rho _{eff}$, $w_{eff}$, $a$ can easily be
obtained using equations (\ref{i}) and (\ref{k}) (or (\ref{l})). Hence, we
can obtain scalar function $\phi (t)$ using equation (\ref{H}) and
eliminating $t$ from $\phi (t)$ and using in (\ref{G}), we can obtain the
potential function $V(\phi )$ for any model resulting from the
parametrization of $q(t)$, $H(t)$ or $a(t)$. It is to be noted that for
quintessence field ($\omega =+1$), from equation (\ref{H}) we can have $\phi
(t)=\phi _{i}+\int \left[ \left( 1+w_{eff}\right) \rho _{eff}-\rho _{0}a^{-3}%
\right] ^{\frac{1}{2}}dt$ while for phantom field ($\omega =-1$), we can
write the scalar function $\phi (t)=\phi _{i}-\int \left[ \left(
1+w_{eff}\right) \rho _{eff}-\rho _{0}a^{-3}\right] ^{\frac{1}{2}}dt$.

\subsection{Potential in $q(t)$ parametrized model}

Equations (\ref{G}) and (\ref{H}) can be written as a single unknown
variable $q(t)$ as

\begin{equation}
V(\phi )=\frac{\left( 2-q\right) M_{p}^{2}}{\left\{ q_{0}+\int \left(
1+q\right) dt\right\} ^{2}}-\frac{\rho _{0}}{2a_{0}^{3}}\exp \left\{ -3\int 
\frac{dt}{q_{0}+\int \left( 1+q\right) dt}\right\} \text{,}  \label{V-DP}
\end{equation}%
where $q_{0}$ and $a_{0}$ are integrating constants. The scalar function $%
\phi (t)$ is given by 
\begin{equation}
\omega \phi (t)=\phi _{i}+\int \left[ \frac{2\left( 1+q\right) M_{p}^{2}}{%
\left\{ q_{0}+\int \left( 1+q\right) dt\right\} ^{2}}-\frac{\rho _{0}}{%
a_{0}^{3}}\exp \left\{ -3\int \frac{dt}{q_{0}+\int \left( 1+q\right) dt}%
\right\} \right] ^{\frac{1}{2}}dt\text{.}  \label{PHI-DP}
\end{equation}

The potential for the Berman's parametrization {\small \cite{Berman}} of
constant deceleration parameter $q(t)=m-1$, is then obtained as

\begin{equation}
V(\phi )=\frac{1}{2}\left[ \frac{\left( 3-\beta \right) M_{p}^{2}}{\left(
q_{0}+\beta t\right) ^{2}}-\frac{\rho _{0}}{a_{0}^{3}\left( q_{0}+\beta
t\right) ^{\frac{3}{\beta }}}\right]  \label{EX-DP}
\end{equation}%
together with%
\begin{equation}
\omega \phi (t)=\phi _{i}+\int \left[ \frac{2mM_{p}^{2}}{\left(
q_{0}+mt\right) ^{2}}-\frac{\rho _{0}}{a_{0}^{3}\left( q_{0}+mt\right) ^{%
\frac{3}{m}}}\right] ^{\frac{1}{2}}dt\text{.}  \label{EX-DP-phi}
\end{equation}%
At late times, when the dark energy overtakes the matter energy i.e. $\rho
_{eff}=\rho _{\phi }$, we have $\omega \phi (t)-\phi _{i}=\sqrt{\frac{2}{m}}%
M_{p}\ln \left( q_{0}+mt\right) $ and the potential is found to an
exponential potential in the form%
\begin{equation}
V(\phi )=\left( 3-2m\right) M_{p}^{2}\exp \left\{ -\frac{\sqrt{2m}}{M_{p}}%
\left( \omega \phi -\phi _{i}\right) \right\} \text{.}  \label{BM-V-N}
\end{equation}%
Similarly, the potential for Linearly varying deceleration parameter model
(LVDP) {\small \cite{akarsu}} $q(t)=-2\alpha t+\beta -1$ (at late times) is
given as 
\begin{equation}
V(t)=\frac{\left( 3-\beta +2\alpha t\right) M_{p}^{2}}{\left( q_{0}+\beta
t-\alpha t^{2}\right) ^{2}},  \label{EX-DP-V}
\end{equation}%
where $t$ is to be eliminated from%
\begin{equation}
\frac{\left( 4q_{0}\alpha +\beta ^{2}\right) ^{\frac{1}{4}}}{2\sqrt{2}M_{p}}%
(\omega \phi -\phi _{i})=\tan ^{-1}\frac{\sqrt{\beta -2\alpha t}}{\left(
4q_{0}\alpha +\beta ^{2}\right) ^{\frac{1}{4}}}-\tanh ^{-1}\frac{\sqrt{\beta
-2\alpha t}}{\left( 4q_{0}\alpha +\beta ^{2}\right) ^{\frac{1}{4}}}.
\label{EX-DP-phi-LVDP}
\end{equation}

\subsection{Potential in $a(t)$ parametrized model}

Equations (\ref{G}) and (\ref{H}) can be written as a single unknown
variable $a(t)$ as

\begin{equation}
V(\phi )=M_{p}^{2}\left( 2+\frac{a\ddot{a}}{\dot{a}^{2}}\right) \frac{\dot{a}%
^{2}}{a^{2}}-\frac{\rho _{0}}{2a^{3}}  \label{V-SF}
\end{equation}%
together with the scalar function 
\begin{equation}
\omega \phi (t)=\phi _{i}+\int \left[ 2M_{p}^{2}\left( 1-\frac{a\ddot{a}}{%
\dot{a}^{2}}\right) \frac{\dot{a}^{2}}{a^{2}}-\frac{\rho _{0}}{a^{3}}\right]
^{\frac{1}{2}}dt\text{.}  \label{PHI-SF}
\end{equation}

The potential for the power law cosmology {\small \cite{PLC}} $a(t)=\beta
t^{n}$, is given by

\begin{equation}
V(\phi )=M_{p}^{2}\frac{n(3n-1)}{t^{2}}-\frac{\rho _{0}}{2\beta ^{3}t^{3n}}
\label{EX-SF}
\end{equation}%
together with%
\begin{equation}
\omega \phi (t)=\phi _{i}+\int \left[ 2M_{p}^{2}\frac{n}{t^{2}}-\frac{\rho
_{0}}{\beta ^{3}t^{3n}}\right] ^{\frac{1}{2}}dt\text{.}  \label{EX-SF-phi}
\end{equation}%
At late times, when the dark energy overtakes the matter energy i.e. $\rho
_{eff}=\rho _{\phi }$, we have $\omega \phi (t)-\phi _{i}=\sqrt{2n}M_{p}\ln
t $ and the potential is found to be again an exponential potential in the
form%
\begin{equation}
V(\phi )=n(3n-1)M_{p}^{2}\exp \left\{ -\sqrt{\frac{2}{n}}\frac{1}{M_{p}}%
\left( \omega \phi -\phi _{i}\right) \right\} \text{.}  \label{V-PLC}
\end{equation}

\subsection{Potential in $H(t)$ parametrized model}

Equations (\ref{G}) and (\ref{H}) can be written as a single unknown
variable $H(t)$ as

\begin{equation}
V(\phi )=M_{p}^{2}\left[ 3H^{2}+\dot{H}\right] -\frac{\rho _{0}}{2a_{0}^{3}}%
\exp \left\{ -3\int H(t)dt\right\}  \label{V-HP}
\end{equation}%
together with the expression of scalar function 
\begin{equation}
\omega \phi (t)=\phi _{i}+\int \left[ -2M_{p}^{2}\dot{H}-\frac{\rho _{0}}{%
a_{0}^{3}}\exp \left\{ -3\int H(t)dt\right\} \right] ^{\frac{1}{2}}dt\text{.}
\label{PHI-HP}
\end{equation}

The potential for the parametrized Hubble function of the form $H(t)=\frac{%
\beta t^{m}}{\left( t^{n}+\alpha \right) ^{p}}$ {\small \cite{SKJP},} is
found to be

\begin{equation}
V(\phi )=M_{p}^{2}\left[ \frac{3\beta ^{2}t^{2m}}{\left( t^{n}+\alpha
\right) ^{2p}}+\beta \left( \frac{mt^{m-1}}{\left( t^{n}+\alpha \right) ^{p}}%
-\frac{npt^{m+n-1}}{\left( t^{n}+\alpha \right) ^{p+1}}\right) \right] -%
\frac{\rho _{0}}{2a_{0}^{3}}\exp \left\{ -3\beta \int \frac{t^{m}}{\left(
t^{n}+\alpha \right) ^{p}}dt\right\}  \label{EX-HP}
\end{equation}%
together with%
\begin{equation}
\omega \phi (t)=\phi _{i}+\int \left[ -2M_{p}^{2}\beta \left( \frac{mt^{m-1}%
}{\left( t^{n}+\alpha \right) ^{p}}-\frac{npt^{m+n-1}}{\left( t^{n}+\alpha
\right) ^{p+1}}\right) -\frac{\rho _{0}}{a_{0}^{3}}\exp \left\{ -3\beta \int 
\frac{t^{m}}{\left( t^{n}+\alpha \right) ^{p}}dt\right\} \right] ^{\frac{1}{2%
}}dt.  \label{SK-H}
\end{equation}%
For a specific model with $m=0,$ $n=1,$ $p=\frac{1}{2}$ (Model-VI of {\small 
\cite{SKJP})}, we have $H(t)=\frac{\beta }{\sqrt{t+\alpha }}$. At late
times, when the dark energy overtakes the matter energy i.e. $\rho
_{eff}=\rho _{\phi }$, we have $\omega \phi (t)-\phi _{i}=4M_{p}\sqrt{\beta }%
\left( t+\alpha \right) ^{\frac{1}{4}}$ and the potential is obtained as%
\begin{equation}
V(\phi )=\frac{256M_{p}^{3}\beta ^{\frac{3}{2}}}{\left( \omega \phi -\phi
_{i}\right) }\left[ 3\beta -\frac{8\left( M_{p}\sqrt{\beta }\right) ^{\frac{1%
}{2}}}{\sqrt{\omega \phi -\phi _{i}}}\right] .  \label{M6-SK}
\end{equation}

\section{Potential for Tachyonic field}

We consider an action describing a general tachyon field $\phi $ as

\begin{equation}
S=-\int d^{4}xV(\phi )\sqrt{-\det \left( g_{\mu \nu }+\partial _{\mu }\phi
\partial ^{\mu }\phi \right) }\text{,}  \label{TA}
\end{equation}%
where $V(\phi )$ is the potential function for the tachyon field. In the
flat FRW background the energy density $\rho _{\phi }$ and pressure $p_{\phi
}$ of the tachyon field can be written as

\begin{equation}
\rho _{\phi }=\frac{V\left( \phi \right) }{\sqrt{1-\dot{\phi}^{2}}},
\label{Ta}
\end{equation}

\begin{subequations}
\begin{equation}
p_{\phi }=-V\left( \phi \right) \sqrt{1-\dot{\phi}^{2}}\text{.}  \label{Tb}
\end{equation}

Here also, we consider two fluid (tachyons and matter) model. If we assume
the minimal interaction between matter field and tachyon field and making
use of the Friedmann equations (\ref{i}) and (\ref{j}) along with the
perfect fluid equation of state, we obtain the tachyonic potential as 
\end{subequations}
\begin{equation}
V(\phi )=\sqrt{-w_{eff}\rho _{eff}\left( \rho _{eff}-\rho _{0}a^{-3}\right) }
\label{Tc}
\end{equation}

and

\begin{equation}
\phi (t)-\phi _{i}=\int \sqrt{\frac{\left( 1+w_{eff}\right) \rho _{eff}-\rho
_{0}a^{-3}}{\left( \rho _{eff}-\rho _{0}a^{-3}\right) }}dt\text{, }\phi _{i}%
\text{ is an integrating constant.}  \label{Td}
\end{equation}%
\qquad \qquad

As in the case of quintessence and phantom fields, we can obtain the tachyon
potential $V(\phi )$ and the tachyon field $\phi (t)$ using the relation (%
\ref{Tc}) and (\ref{Td}) for any parametrization of any cosmological
parameter $a(t)$, $q(t)$, $H(t)$ where the quantities $\rho _{eff}$, $%
w_{eff} $, $a$ can easily be obtained using equations (\ref{i}) and (\ref{k}%
) (or (\ref{l})).

Tachyonic potential for power law cosmology {\small \cite{PLC}} $a(t)=\beta
t^{n}$, is obtained as

\begin{equation}
V(\phi )=M_{p}\sqrt{\frac{(3n^{2}-2n)}{t^{2}}\left( \frac{3n^{2}M_{p}^{2}}{%
t^{2}}-\frac{\rho _{0}}{\beta ^{3}t^{3n}}\right) }  \label{TV-PLC}
\end{equation}%
together with%
\begin{equation}
\phi (t)-\phi _{i}=\int \left[ \left( \frac{2nM_{p}^{2}}{t^{2}}-\frac{\rho
_{0}}{\beta ^{3}t^{3n}}\right) /\left( \frac{3n^{2}M_{p}^{2}}{t^{2}}-\frac{%
\rho _{0}}{\beta ^{3}t^{3n}}\right) \right] ^{\frac{1}{2}}dt\text{.}
\label{Tach-phi}
\end{equation}%
At late times, when $\rho _{eff}=\rho _{\phi }$, we have $\phi -\phi _{i}=%
\sqrt{\frac{2}{3n}}t$ and the potential%
\begin{equation}
V(\phi )=2M_{p}^{2}\sqrt{n^{2}-\frac{2}{3}n}\frac{1}{\left( \phi -\phi
_{i}\right) ^{2}}\text{.}  \label{Tach-BM-A}
\end{equation}

Tachyonic potential for Berman's model of constant deceleration parameter 
{\small \cite{Berman}} $q(t)=m-1$, is given by

\begin{equation}
V(\phi )=\frac{\sqrt{3-2m}M_{p}}{\left( q_{0}+mt\right) }\sqrt{\frac{%
3M_{p}^{2}}{\left( q_{0}+mt\right) ^{2}}-\frac{\rho _{0}}{a_{0}^{3}\left(
q_{0}+mt\right) ^{\frac{3}{m}}}}  \label{TV-Berman}
\end{equation}%
together with%
\begin{equation}
\phi (t)-\phi _{i}=\int \left[ \left( \frac{2mM_{p}^{2}}{\left(
q_{0}+mt\right) ^{2}}-\frac{\rho _{0}}{a_{0}^{3}\left( q_{0}+mt\right) ^{3/m}%
}\right) /\left( \frac{3M_{p}^{2}}{\left( q_{0}+mt\right) ^{2}}-\frac{\rho
_{0}}{a_{0}^{3}\left( q_{0}+mt\right) ^{3/m}}\right) \right] ^{\frac{1}{2}%
}dt.  \label{Tach-BM-phi}
\end{equation}%
At late times, when $\rho _{eff}=\rho _{\phi }$, we have $\phi -\phi _{i}=%
\sqrt{\frac{2m}{3}}t$ and the potential is given as%
\begin{equation}
V(\phi )=\frac{\sqrt{3\left( 3-2m\right) }M_{p}^{2}}{\left\{ q_{0}+\sqrt{%
\frac{3m}{2}}\left( \phi -\phi _{i}\right) \right\} ^{2}}.  \label{Tach-BM-B}
\end{equation}%
Similarly, the potential for LVDP model {\small \cite{akarsu}} $%
q(t)=-2\alpha t+\beta -1$, is given by%
\begin{equation}
V(t)=\sqrt{3}M_{p}^{2}\frac{\sqrt{3+2\left( 2\alpha t-\beta \right) }}{%
\left( q_{0}+\beta t-\alpha t^{2}\right) ^{2}},  \label{Tach-LVDP}
\end{equation}%
where $t$ is to be eliminated from $\phi (t)=\phi _{i}-\frac{\sqrt{6}}{%
9\alpha }\left( \beta -2\alpha t\right) ^{3/2}$.

Tachyonic potential for the $H(t)$ parametrized model {\small \cite{SKJP}} $%
H(t)=\frac{\beta }{\sqrt{t+\alpha }}$ (Model-VI in {\small \cite{SKJP}}) is
obtained as

\begin{equation}
V(\phi )=M_{p}\sqrt{\beta }\sqrt{\frac{\left( 3\beta \sqrt{t+\alpha }%
-1\right) }{\left( t+\alpha \right) ^{\frac{3}{2}}}\left( \frac{3\beta
^{2}M_{p}^{2}}{t+\alpha }-\frac{\rho _{0}}{a_{0}^{3}\exp (6\beta \sqrt{%
t+\alpha })}\right) }  \label{TV-Hub}
\end{equation}%
together with%
\begin{equation}
\phi (t)-\phi _{i}=\int \left[ \left( \frac{\beta M_{p}^{2}}{\left( t+\alpha
\right) ^{3/2}}-\frac{\rho _{0}}{a_{0}^{3}\exp (6\beta \sqrt{t+\alpha }}%
\right) /\left( \frac{3\beta ^{2}M_{p}^{2}}{t+\alpha }-\frac{\rho _{0}}{%
a_{0}^{3}\exp (6\beta \sqrt{t+\alpha }}\right) \right] ^{\frac{1}{2}}dt.
\label{Tach-H-m-6}
\end{equation}%
At late times, when $\rho _{eff}=\rho _{\phi }$, we have $\phi -\phi _{i}=%
\frac{4}{3\sqrt{3\beta }}\left( t+\alpha \right) ^{3/4}$ and the potential
is given as%
\begin{equation}
V(\phi )=\frac{8}{9}\frac{1}{2^{1/3}}\frac{1}{\beta ^{1/6}}M_{p}^{2}\left[ 
\frac{9\beta ^{4/3}}{\left( \phi -\phi _{i}\right) ^{8/3}}-\frac{2^{2/3}}{%
\left( \phi -\phi _{i}\right) ^{10/3}}\right] ^{\frac{1}{2}}.
\label{Tach-V-m6-sk}
\end{equation}%
Following the same procedure, scalar field potentials cab be constructed
either explicitly or implicitly for any cosmological parametrization.

\section{Conclusion}

\qquad \qquad In this paper, we consider models based upon a specific scheme
of parametrization. We have constructed the scalar field potentials in $q(t)$%
, $a(t)$ and $H(t)$ parametrized models for quintessence, phantom and
tachyonic fields in the FRW framework. In case of constant deceleration
parameter or power law cosmology, the scalar field potential reduces to
exponential form as expected. In case of tachyon field, the potential
corresponding to scaling solution is provided by inverse power law, $V(\phi
)\sim \phi ^{-2}$ as noted earlier. For a specific model (model-VI in 
{\small \cite{SKJP}) }resulting from a parametrization of $H$, the potential 
$V(\phi )\sim \left[ V_{1}(\phi )+V_{1}(\phi )\right] $ where $V_{1}(\phi
)\sim \phi ^{-1}$ and $V_{1}(\phi )\sim \phi ^{-3/2}$ for (non)phantom case
and $V(\phi )\sim \left[ V_{3}(\phi )+V_{4}(\phi )\right] $ where $%
V_{3}(\phi )\sim \phi ^{-8/3}$ and $V_{1}(\phi )\sim \phi ^{-10/3}$ in case
of tachyon. Similarly, we can also construct the scalar field potentials for
all other $H(t)$ parametrized models obtained in {\small \cite{SKJP}}. The
potentials for the linearly varying deceleration parameter model have also
been obtained for both (non)phantom and tachyonic fields as implicit
functions of $\phi $ and\ $t$. In principle, for any scheme of
parametrization of $a(t)$, $q(t)$, $H(t)$, $w(t)$, $\rho (t)$, the scalar
field potentials for quintessence, phantom and tachyonic fields can be
constructed.

\bigskip

{\Large Acknowledgement} \textit{The authors wish to thank M. Sami for his
useful comments and suggestions throughout the work. The authors also thank to S. D. Odintsov for his valuable comments. Author SKJP wish to
thank Department of Atomic Energy (DAE), Government of India for financial
support through the post-doctoral fellowship of National Board of Higher
Mathematics (NBHM).}

\end{document}